\documentstyle[aps,epsf]{revtex}

\newcommand{\be}{\begin{eqnarray}}
\newcommand{\ee}{\end{eqnarray}}
\newcommand{\beq}{\begin{equation}}
\newcommand{\eeq}{\end{equation}}
\newcommand{\ba}{\begin{array}}
\newcommand{\ea}{\end{array}}
\begin{document}
\thispagestyle{empty}
\begin{flushright}
JLAB-THY-00-29 \\
RUB-TPII-14/00 \\
\end{flushright}
\vspace{1cm}
\begin{center}
{\bf\Large 
 DVCS amplitude with  kinematical twist-3 terms} \\[1cm]
\end{center}
\begin{center}
A.V. RADYUSHKIN$^{a,b,1}$,
 C. WEISS$^{c,2}$
\\[2mm]
{\em $^a$Physics Department, Old Dominion University,} \\
{\em Norfolk, VA 23529, USA}
\\[2mm] 
{\em $^b$Theory Group, Jefferson Lab,} \\
{\em Newport News, VA 23606, USA}
\\[2mm] 
{\em $^c$ Institut f\"ur Theoretische Physik II,
Ruhr-Universit{\"a}t  Bochum, Germany} 
\end{center}
\vspace{1cm}
\begin{abstract}
We compute the amplitude of deeply virtual Compton scattering 
(DVCS) 
 using the calculus 
of QCD string operators in coordinate representation. To restore 
the electromagnetic gauge invariance (transversality) of the 
twist-2 amplitude 
 we include the   operators  of twist-3  
which appear as total derivatives of twist-2 operators.
Our results are equivalent to a
Wandzura-Wilczek approximation for 
twist-3 skewed parton distributions. 
We find that, for the tensor amplitude
$T_{\mu\nu}$,  this approximation 
gives a finite result for the  
longitudinal polarization  of the virtual photon
and a divergent result in case of 
 transverse polarization. 
 However, the divergent part
 has zero projection
 onto the polarization
 vector of the final real photon.

\end{abstract}
\vfill
\rule{5cm}{.2mm} \\
{\footnotesize $^{\rm 1}$ E-mail: radyush@jlab.org; also 
at Laboratory of Theoretical Physics, JINR, Dubna, Russian Federation} \\
{\footnotesize $^{\rm 2}$ E-mail: weiss@tp2.ruhr-uni-bochum.de} \\

\section{Introduction}

The deeply virtual Compton scattering (DVCS) 
process, in which a highly virtual
photon $\gamma^*(q_1)$ produces a real photon $\gamma(q_2=q_1+r)$
at small invariant momentum transfer 
$t= r^2$,  is 
receiving  a lot  
of attention 
as a potential source of new information about  
 the  nucleon structure
in terms of so-called ``skewed'' parton distributions (SPD's) 
\cite{Ji:1997nm,Radyushkin:1997ki,Collins:1997fb,Muller1994}. 
The first experimental observation of DVCS has been already reported 
by the ZEUS collaboration \cite{Saull:1999kt}.
Originally, only the purely   twist--2 contribution to the tensor 
amplitude
$T_{\mu\nu}$  amplitude
   was 
calculated \cite{Ji:1997nm,Radyushkin:1997ki}, which  
 does  not  include 
subleading  terms 
 proportional to the transverse 
part $r_{\perp} \equiv \Delta$ 
of the momentum transfer.
As a result,  the twist-2 amplitude    formally  
  violates the 
   electromagnetic (EM) gauge 
invariance  
in terms linear in  $\Delta$. 
 Guichon and Vanderhaeghen \cite{Guichon} proposed 
to add an {\it ad hoc}  $O(\Delta)$ term  
producing  an ``improved'' expression which is  EM 
gauge invariant  up to $O(\Delta^2)$ terms.   
Their prescription  was  recently supported  
by several groups 
\cite{Anikin:2000em,Penttinen:2000dg,Belitsky:2000vx}
 which derived this term
 in a regular way  as a   kinematical twist-3 contribution.  
 Furthermore,  Anikin {\it et al.}  
\cite{Anikin:2000em}, using    the momentum--space 
collinear expansion, have  obtained expressions 
which include (in case of the pion target) all the relevant 
twist--3   operators.     
The DVCS amplitude to accuracy $1/\sqrt{ -q_1^2}$
was also calculated  by Penttinen
{\it et al.} \cite{Penttinen:2000dg}  in 
a parton model approach.   
 Within the light-cone expansion framework,
 Belitsky and M{\"u}ller \cite{Belitsky:2000vx} 
 analyzed both quark and quark-gluon contributions and 
  demonstrated that, 
to get  a gauge invariant result up to terms 
of order $t/q_1^2$,  it is sufficient to retain only the part of the 
twist--3 SPD's which is obtained by Wandzura--Wilczek (WW)
type formulas from the twist--2 distributions. 
Recently, Kivel {\it et al.} \ \cite{Kivel:2000rb}  established
that the WW-approximation  expression 
for the transverse polarization of the virtual photon diverges.  
The mathematical aspects of twist   decomposition were 
discussed by Bl{\"u}mlein and Robaschik \cite{Blumlein:2000cx}, 
 Geyer and Lazar \cite{Geyer2000}.

Our goal is to analyze the DVCS amplitude within 
the QCD string operator approach of Balitsky 
and Braun \cite{Balitsky:1989bk},
which proved to be a powerful tool to 
investigate the higher-twist effects. 
Here, we consider  only the kinematical
twist-3 terms.  In this sense, our results are equivalent
to the WW approximation.  
In addition to offering an alternative derivation of this approximation, 
we incorporate the formalism of 
 double distributions \cite{Radyushkin:1997ki,Radyushkin:sssdd}
 which provides 
 a simple way of deriving relations
 between new SPD's. For instance, the fact that 
 the WW approximation for the amplitude
 $T_{\mu\nu}$ 
 gives finite results  for longitudinally 
polarized photon, but diverges in the case of transverse polarization, 
can  be easily understood on the basis of the  formulas relating
these SPD's and the basic twist-2 DD's. 
We also note that the divergent part of $T_{\mu\nu}$ 
has a vanishing projection onto the 
polarization  of the final real photon.

\section{DVCS amplitude}

{\it Generalities.} 
The  virtual Compton scattering amplitude is derived from the correlation 
function
\begin{eqnarray}
T_{\mu\nu}  &=& 
i \int d^4 x \; \int d^4 y \; e^{-i (q_1 x) + i(q_2  y)}
\langle p_2 | {\rm T} \left\{ J_\mu (x) J_\nu (y) \right\}
| p_1 \rangle \ ,
\label{corr}
\end{eqnarray}
where $J^\mu (x)$ is the  electromagnetic  current 
operator. Due to current conservation 
$\partial^\mu J_\mu (x) = 0$, 
this function is transverse with respect to the incoming and 
outgoing photon momenta:
$
q_1^{\mu}\, T_{\mu\nu} =  0 \, , \  
q_2^{\nu} \,T_{\mu\nu}  =  0 \, . 
$
It is convenient to switch to   symmetric  variables
$q= (q_1+q_2)/2$, $r=q_2-q_1$ and $p=(p_1+p_2)/2$.  Then,  the
transversality conditions  convert 
into two  relations 
\begin{eqnarray} 
             q^{\mu}T_{\{\mu\nu\}}= 
            {r^{\mu}\over 2} T_{[\mu\nu]} \   \   ,  \   \
               q^{\mu}T_{[\mu\nu]}= 
            {r^{\mu}\over 2} T_{\{\mu\nu\}}
	    \label{trans} 
\end{eqnarray}
connecting the 
symmetric $T_{\{\mu\nu\}}\equiv (T_{\mu\nu}+ T_{\nu\mu})/2$
and  antisymmetric 
$T_{[\mu\nu]} \equiv (T_{\mu\nu}- T_{\nu\mu})/2$
parts of $T^{\mu\nu}$.
In the $r=0$ forward limit,   the two relations
decouple  to give  the DIS 
transversality conditions $q^{\mu} T_{\{\mu\nu\}}=0$,
$q^{\mu} T_{[\mu\nu]}=0$. 

In DVCS, 
 the initial 
photon is in the Bjorken  kinematics \{$ - q_1^2 \rightarrow \infty, 
(p_1  q_1) \rightarrow \infty, 
 x_B  \equiv -q_1^2 / [2 (p_1  q_1)]$ fixed\}  and the 
 final one is 
 real $q_2^2 = 0$. Since $q_2^2 = q_1^2 +2 (q_1r) +t$,
 the momentum transfer $r$ in this process 
should have  a large 
component in the direction of $p$,  
with $(r  q_1)$ close to $- q_1^2/2$ for small $t$.  
The size of this component is characterized 
by the skewedness parameter $\eta  \equiv  (r q)/2 (p q)$.
For DVCS, $\eta$  coincides, up to $O(t)$ terms,  with 
the generalized Bjorken variable 
$
\xi  \equiv {-q^2}/{2 (p  q)}\, . 
$ 
Hence, the momentum transfer may  be split into 
the component parallel to $p$  
and the remainder $\Delta$ 
\be 
r= 2 \xi p  +  \Delta \ , 
\label{Delta}
\ee  
which  in the $t=0$, $p^2=0$  limit is  transverse both to
$p$ and $q$:  
$(\Delta  q) = -t/4$,  $(\Delta  p)  = -2 \xi p^2$.

 {\it Coordinate representation.} An  efficient  way  to study 
the behavior of Compton amplitudes
in the Bjorken limit is to use the light--cone expansion 
for the 
product $
\Pi_{\mu\nu} (x, y)  \equiv  i {\rm T} J_{\mu}(x) J_{\nu} (y)  $ 
of two vector currents in the coordinate representation.   
Following Balitsky and Braun \cite{Balitsky:1989bk}, 
we start from the formal 
light--cone expansion  in terms of QCD string operators (with
gauge links along the straight line between the fields which, for brevity,
we do  not write  explicitly).  
The leading light--cone singularity is contained 
in the ``handbag'' contribution    
\begin{eqnarray} 
 \Pi_{\mu\nu} (z\,|\, X) 
 =   \frac{4i z_\rho}{\pi^2 z^4} \biggl\{ s_{\mu\rho\nu\sigma} 
 {\cal O}_{\sigma} (z\,|\,X) - 
 \epsilon_{\mu\rho\nu\sigma}  
{\cal O}_{5\sigma} (z\,|\,X)
\biggl \} \  , 
\label{handbag_string}
\end{eqnarray}
where
$s_{\mu\rho\nu\sigma} = g_{\mu\rho} g_{\nu\sigma} - g_{\mu\nu} g_{\rho\sigma}
+ g_{\mu\sigma} g_{\nu\rho} $, \,  $X  = (x + y)/2$,  
$z  =  y - x $,  and 
\be
{\cal O}_{\sigma} (z\,|\,X) &=& \frac1{2i} \,
\left[ \bar\psi (X-z/2) \gamma_\sigma 
\psi (X+z/2) \; - \; (z \rightarrow -z) \right] \  ,
\nonumber \\
{\cal O}_{5\sigma} (z\,|\,X) &=& \frac1{2} \,
\left[ \bar\psi (X-z/2) \gamma_\sigma \gamma_5
\psi (X+z/2) \; + \; (z \rightarrow -z) \right] \  .
\ee

 {\it Twist--2 part.}
The string operators in Eq. (\ref{handbag_string}) do not have a 
definite twist.   
The   twist--2 part  is defined 
by formally Taylor--expanding the string operators in 
Eq. (\ref{handbag_string}) 
in the relative coordinate
 $z$ and retaining only the totally symmetric traceless 
parts of the coefficients in the expansion:
\be
\left[ \bar\psi (X - z/2) \gamma_\sigma \psi (X + z/2) \right]^
{\rm twist-2}
\equiv 
\sum_{n = 0}^\infty \frac{1}{ n!} \;
z_{\mu_1} \ldots z_{\mu_n} \;
\bar\psi (X) \left[
\gamma_{\left\{ \sigma \right. } \stackrel{\leftrightarrow}{D}_{\mu_1}
\ldots \stackrel{\leftrightarrow}{D}_{\left. \mu_n \right\}} 
- \mbox{traces} \right] \psi (X) , 
\label{gamma_LT}
\ee
and similarly for the operator with  $\gamma_\sigma \gamma_5$. 
As shown in Ref.\cite{Balitsky:1989bk}, 
 ``symmetrization'' and ``subtraction of traces''
can be carried out directly at the level of non-local operators.
The part of the string operator corresponding to totally symmetric
local tensor operators is projected out by 
\beq
\left[ \bar\psi (X - z/2) \gamma_\sigma \psi (X + z/2) \right]^{\rm sym}
\;\; = \;\; 
\frac{\partial}{\partial z_\sigma} \int_0^1 dt \; 
\bar\psi (X - t z/2) \hat{z} \psi (X + t z/2) \ 
\label{string_sym}
\eeq
(we use the notation $\hat z \equiv z^\sigma \gamma_\sigma$).
The subtraction of traces in  the local operators implies that 
the twist--2 string operator contracted 
with $z_\sigma$ should satisfy the d'Alembert 
equation with respect to $z$:
\be
\Box_z  \left[ \bar\psi (X - t z/2) \hat{z} \psi (X + t z/2) 
\right]^{\rm twist-2}
= 0.
\label{harmonic_O} \ee

{\it Transversality and twist--3 operators. }  
In the coordinates $X$ and $z$, the transversality conditions (\ref{trans})
 are 
\be
\frac{\partial}{\partial z_\mu} \Pi_{\{\mu\nu\}} (z|X) 
= \frac{1}{2} \frac{\partial}{\partial X_\mu} \Pi_{[\mu\nu ]} (z|X)
 \  \  ,  \  \ 
\frac{\partial}{\partial z_\mu} \Pi_{[\mu\nu ]} (z|X)
= \frac{1}{2} \frac{\partial}{\partial X_\mu} 
\Pi_{\{\mu\nu\}} (z|X) \  . 
\label{transversality_antisym}
\ee
  Consider the part of the current product  given  by 
Eq. (\ref{handbag_string})  with the string operators replaced by 
their twist--2 parts. 
From  Eq. (\ref{harmonic_O})  and 
$({\partial}/{\partial z_\rho})  [ {z_\rho}/{(2\pi^2 z^4)} ]
 = -i \delta^{(4)} (z)$ 
it  follows  that  
$({\partial}/{\partial z_\mu})\, \Pi_{\{\mu\nu\}}^{\rm twist-2}=  0$,   
$({\partial}/{\partial z_\mu}) \,\Pi_{[\mu\nu ]}^{\rm twist-2} 
 = 0.$ 
 Since forward matrix elements  are zero 
for  all total derivative operators,  
 this  guarantees the 
transversality of the  twist--2 contribution
in  the case of deep inelastic scattering.     
In the  non-forward case,  we have 
\mbox{$({\partial}/{\partial X_\mu}) \Pi_{\{\mu\nu\}}^{\rm twist-2}
 \neq  0, \ 
({\partial}/{\partial X_\mu})  \Pi_{[\mu\nu ]}^{\rm twist-2}
 \neq  0\ , 
$} and (\ref{transversality_antisym})  is violated. 
 The non-transverse terms 
in the twist--2 contribution can   only be 
compensated by contributions from 
operators of higher twist.
In fact, the necessary 
operators 
are contained in the   part of the 
string operator which was dropped in taking the twist--2 part.  
Incorporating   QCD equations of motion, it is possible 
 to show \cite{Balitsky:1989bk} that the twist$>2$  part  
 involves  the total derivatives  
 of string operators 
\be
\bar\psi (-z/2) \gamma_\alpha \psi (z/2)
\; - \; \left[ \bar\psi (-z/2) \gamma_\alpha \psi (z/2) \right]^{\rm sym} 
=  \frac{i}{2} \epsilon_{\alpha\xi\rho\kappa} z_\xi
\frac{\partial}{\partial X_\rho} \int_0^1 dt \, t \;
\bar\psi (-tz/2) \gamma_\kappa \gamma_5  \psi (tz/2) +\ldots  \   \   .
\label{rest_total_derivative} 
\ee
The ellipses stand for 
quark--gluon operators  (we do not write them
explicitly since  they  are not needed to  restore transversality of the 
twist--2 contribution).  
The   relation  for the operator with Dirac matrix 
$\gamma_\alpha\gamma_5$ is obtained by changing  
$\gamma_\alpha \to \gamma_\alpha \gamma_5$,
$\gamma_\kappa \gamma_5 \to \gamma_\kappa $.  
The operators 
appearing under the total derivative 
on the R.H.S. of Eq. (\ref{rest_total_derivative}) 
and its $\gamma_\alpha\gamma_5$ analog are  still the  
 full string operators  with 
  no definite twist. Hence, one  can   decompose them 
into a symmetric ({\it i.e.}, twist--2) part and total derivatives,
and so on; thus expressing  the original string operator 
as the sum of its symmetric part and an infinite series of  
arbitrary order total derivatives  
of symmetric operators. 
This series can be summed up in a closed form
(the details of our  calculation 
are   presented elsewhere \cite{rwlong};  
similar expressions were derived independently in 
\cite{Belitsky:2000vx,Kivel:2000rb}). 
 Up to operators whose matrix elements give $O(t)$ contributions 
  to the Compton amplitude, the result is 
\be
\bar\psi (-z/2) \gamma_\sigma \psi (z/2) 
&=& \int_0^1 dv \; 
\left\{  \cos \left[ \frac{i \bar v}{2} 
\; \left (z  \frac{\partial}{\partial X} \right) \right ] 
\frac{\partial}{\partial z_\sigma}
\; + \; \frac{i v}{2}
\sin \left[ \frac{i \bar v}{2} 
\; \left (z  \frac{\partial}{\partial X} \right) \right ]
\frac{\partial}{\partial X_\sigma}
\right\} \bar\psi (-vz/2) \hat z \psi (vz/2) 
\nonumber \\
&&  + \; 
\frac{i}{2} \epsilon_{\sigma\alpha\beta\gamma} z_\alpha 
\frac{\partial}{\partial X_\beta} \frac{\partial}{\partial z_\gamma}
\int_0^1 dv \; \; \int_v^1 du 
\cos \left[ \frac{i \bar u}{2} 
\; \left (z  \frac{\partial}{\partial X} \right) \right ] 
\; \bar\psi (-vz/2) \hat z \gamma_5 \psi (vz/2) 
 + \ldots \ . 
\label{string_deconstructed_scalar}
\ee
An analogous   formula applies to the operators with
$\gamma_\sigma \rightarrow \gamma_\sigma\gamma_5$;  one 
should just replace $\hat z \rightarrow \hat z \gamma_5 , \; \hat z \gamma_5 
\rightarrow \hat z$. 

\section{Parametrization  of nonforward matrix elements}
\label{subsec_spectral}
 {\it Double distributions.} To get 
the amplitude for deeply virtual Compton scattering off a hadronic 
target we need  parametrizations of the hadronic 
matrix elements of the uncontracted 
twist--2 string operators ${\cal O}_{\sigma},{\cal O}_{5\sigma}$
appearing 
in  Eq. (\ref{handbag_string}). 
We will derive them from Eq. (\ref{string_deconstructed_scalar}).   
For simplicity, we  consider  here one quark flavor and  the  pion target,
which has zero spin and practically vanishing mass. 
In this  case, the matrix element of 
the contracted {\it axial}  operator 
$z^\sigma {\cal O}_{5\sigma}(z\,|\,0)$ 
(parametrized in the forward limit by the polarized parton density) 
is identically 
zero.  Thus we need only the 
parametrization for the matrix element 
$\langle p - r/2 \,|\, {\cal O}(z\,|\,0)\, |\, p + r/2 \rangle$
of the contracted {\it vector}
   operator ${\cal O}(z\,|\,0) 
   \equiv z^\sigma {\cal O}_{\sigma}(z\,|\,0)$. 
With respect to $z$,  it can be regarded as a 
function of three invariants $(pz), (rz)$ and $z^2$. 
For dimensional reasons, 
the dependence on $z^2$ is through 
the combinations $t z^2$ and $p^2 z^2$ only. 
Since we are going to  drop $O(t)$ and $O(p^2)$ terms 
 in the Compton amplitude, 
 we  ignore the dependence on $z^2$ and 
 treat this matrix element as a  function
of just two variables $(pz)$ and $(rz)$.
Incorporating the spectral properties
of nonforward matrix elements \cite{Radyushkin:sssdd} , 
we  write   the plane wave expansion in the form 
\begin{eqnarray} 
\langle p - r/2 \,|\, {\cal O}(z\,|\,0)\, | \, p + r/2 \rangle 
 =   2(pz) \, \int_{-1}^1 d\tilde x  
\int_{-1 + |\tilde x|}^{1-|\tilde x|} e^{-i (kz)} 
 f(\tilde x,\alpha) \, d \alpha  
 +  (rz) \, \int_{-1}^1 e^{-i \alpha (rz)/2}\, 
D (\alpha) \,  d \alpha    \  , 
\label{para1}  
\end{eqnarray} 
where 
$k =  \tilde x p +  \alpha r /2$, 
$f(\tilde x,\alpha)$ is the  double distribution (DD) and  
 $D (\alpha)$ is the Polyakov-Weiss ({\rm PW}) distribution amplitude
 \cite{PW} 
absorbing the $(pz)$-independent terms.    
From this parametrization,   we can obtain the matrix elements
of   original uncontracted  string operators, 
 (\ref{string_deconstructed_scalar}),    including  the kinematical 
twist--3 contributions. 
We consider first the part coming from the double distribution 
term in Eq. (\ref{para1}); the contributions from the PW--term will 
be included separately. 
In matrix elements, the total derivative turns into the momentum
transfer, $
i{\partial}/{\partial X_{\sigma}} \rightarrow 
r_{\sigma}=  2 \xi p_{\sigma} + \Delta_{\sigma}
$.
Similarly, we  write $k=(\tilde x+\xi \alpha)p + {\alpha} \Delta / 2$.
This gives   
\begin{eqnarray} 
 \frac{1}{2} \, 
\langle p - r/2\, |\, {\cal O}_\sigma (z\,|\,0)\, | \, p +r/2  \rangle 
=  \int_{-1}^1 d \tilde x \int_{-1 + |\tilde x|}^{1-|\tilde x|} 
d \alpha \, f(\tilde x,\alpha)  \, \biggl \{p_{\sigma}
e^{-i (\tilde x+ \xi \alpha )(pz) -i\alpha (\Delta z)/2 }  
\nonumber \\ 
+ \frac12\, \bigl [\Delta_{\sigma} (pz) - p_{\sigma}(\Delta z)\bigr ]
\int_0^1 dv \, v\, e^{-i v (\tilde x + \xi \alpha )(pz)-iv\alpha (\Delta z)/2 } 
 \, 
  \bigl [  \sin  (\bar v (rz)/2) 
- i  \alpha  \cos (\bar v   (rz)/2)  \bigr ] \biggr \} \ . 
\label{para5} 
\end{eqnarray}
\par

 {\it Skewed distributions.} Expanding $\exp[-i\alpha (\Delta z)/2]=
 1-i\alpha (\Delta z)/2 +\ldots$ and keeping only terms
 up to  those linear  in  the transverse 
 momentum\footnote{Because of  this truncation, the  
  $\Delta_\mu \Delta_\nu$ terms in the expression
  for the amplitude $T_{\mu\nu}$ will be  lost.
  If needed, they can be kept;  see 
  the discussion after Eq. (\ref{xider}).} 
  $\Delta$ we get an 
 expression in which 
the spectral parameter $\tilde x$ appears in the exponential factors 
 only in the combination 
$
x  \equiv  \tilde x + \xi \alpha $. 
Thus,  we can introduce two skewed parton distributions: 
\begin{equation}
\left.
\begin{array}{r}
H(x, \xi) \\[1.5ex]
A (x, \xi)
\end{array}
\right\}
\;\; \equiv \;\; \int_{-1}^1 d\tilde x  
\int_{-1 + |\tilde x|}^{1-|\tilde x|}  d \alpha   \, 
\delta (x  - \tilde x  -  \xi \alpha ) \, f(\tilde x, \alpha) 
\; \left\{
\begin{array}{r}
1 \\[1.5ex]
\alpha
\end{array}
\right.
\label{reduction}
\end{equation}
Note that,  in  our 
case, the DD  $f(\tilde x, \alpha)$ is 
even in $\alpha$ and odd in $\tilde x$. 
As a result, the  functions $H$ and $A$ 
satisfy the symmetry relations
\be 
H(x , \xi) = -H(-x , \xi) \  \  ,  \  \  H(x , \xi) =   H(x , -\xi)  \  \  ,  \  \ 
A(x , \xi) =  A(-x , \xi)  \  \  ,  \  \   A(x , \xi) = -A(x , -\xi) .
\label{symm}
\ee
Furthermore, because of the antisymmetry of the combination 
$ \alpha f (\tilde x, \alpha)$  
with respect both to
$x$ and $\alpha$ we have
\begin{equation} 
\int_{0}^1 dx \, A (x, \xi) \;\; = \;\; 0 .
\end{equation}
Hence, the distribution $A (x, \xi)$ 
cannot be a  positive-definite function  
on $0 \leq x \leq 1$.

Uniting the cosine and sine functions
with the overall exponential factor, $e^{-i vx (pz)}$,
one gets $vx \pm \bar v \xi$ combinations.
Using  (\ref{symm}), 
one can   arrange that only $vx + \bar v \xi$
would appear: 
\begin{eqnarray} && 
\frac{1}{2}  \, 
\langle p-r/2 \, | \, {\cal O}_{\sigma} (z\, |\, 0) \,| \,p+r/2  \rangle =
p_{\sigma} \int_{-1}^1 d x  \, 
   e^{-i x(pz) } \biggl [ H(x, \xi) 
-  \frac{i (\Delta z)}{2} A(x, \xi) \biggr ]
\nonumber \\ 
&+&\frac{i}{2} \, \bigl [\Delta_{\sigma} (pz)  - 
p_{\sigma}(\Delta z)\bigr ]
\int_{-1}^1 d x \, \bigl [
H (x, \xi)- 
 A(x, \xi)  
 \bigr ]\int_0^1 dv \, v \, 
 \cos [(v x +\bar v \xi) (pz)]  \  .
\label{para7} 
\end{eqnarray}
In a similar fashion, we get parametrization
for  the matrix element of the axial  
string operator (\ref{string_deconstructed_scalar}):
\begin{eqnarray} 
 \frac{1}{2}\,
\langle p-r/2 \, | \, {\cal O}_{5\sigma } (z\,|\,0) \,  |\,  p+r/2 \rangle =  
\frac{i}{2}  \,  \epsilon_{\sigma\alpha\beta\gamma} \, z_\alpha 
 \Delta_{\beta}    p_{\gamma}
\int_{-1}^1 d x \, \bigl [ 
H (x, \xi)- 
 A(x, \xi)  
 \bigr ]
    \int_0^1 dv \, v \, 
  \sin[(v x +\bar v \xi) (pz)] 
\  .  
\label{parax5} 
\end{eqnarray}
 Note that it is expressed in terms of the
same skewed distributions $H(x, \xi)$ and $A(x, \xi)$ which,  
in turn, are determined by the
original double distribution  
$f(\tilde x, \alpha) $, see Eq. (\ref{reduction}). 

\section{ DVCS amplitude for pion target}

 {\it DD-generated contribution.} 
Substituting the parametrizations (\ref{para7}) and (\ref{parax5}) 
into Eq. (\ref{handbag_string})  
and performing the Fourier integral over the separation $z$ we
obtain the  Compton amplitude 
\be
T_{\mu\nu}   & = & \;\; \frac{1}{(pq)} 
 \; 
 \left[ p_\mu q_\nu + q_\mu p_\nu - g_{\mu\nu} (pq) + 2 \xi p_\mu p_\nu
+  \frac{\Delta_\mu}{2} p_\nu- p_\mu\frac{\Delta_\nu}{2}    
\right] \int_{-1}^1 dx \frac{H(x, \xi)}{x - \xi + i0} 
\nonumber
\\ 
&+&   \, 
\frac1{2(pq)} \, \int_{-1}^1 dx \,  R(x, \xi)  
\int_0^1 dv  \,  
\frac{ (q_\mu + 3 \xi p_\mu)\Delta_\nu }{\xi + v x + \bar v \xi - i0 }
 + \frac1{2(pq)} \, \int_{-1}^1 dx \,  R(x, \xi)  
\int_0^1 dv \; \frac{\Delta_\mu (q_\nu + \xi p_\nu) }
{-\xi + v x + \bar v \xi + i0} 
 \  ,
\label{Compton_Anikin}
\ee
where $R (x, \xi)$ is  a new  SPD describing the kinematical
twist-3 contributions: 
\be
 R (x, \xi) \equiv \frac{{\partial H(x, \xi)}}{\partial x}
 - \frac{{\partial A(x, \xi)}}{\partial x} \  . 
 \label{R}
\ee 
This result for  the Compton amplitude contains  the 
same tensor structures 
as those   obtained in Refs.\cite{Anikin:2000em,Penttinen:2000dg}.
All three terms in 
Eq. (\ref{Compton_Anikin})  
are individually transverse up to terms of
order $t,p^2$. 

{\it Singularities.} 
 The 
first term is the  twist--2 part 
with the  tensor structure corrected exactly as suggested by 
Guichon and Vanderhaeghen \cite{Guichon}.
The integral over $x$ exists if  $H(x,\xi)$ is 
continuous at $x = \xi$, which is the case for SPD's derived from 
the DD's which are less singular than $1/\tilde x^2$ for $\tilde x =0$ 
and are continuous otherwise (see \cite{Musatov2000}). 
 In particular, continuous SPD's were  obtained in
model calculations of  SPD's at a low scale in the instanton 
vacuum \cite{chiralbag}. 
The second term contributes only to the helicity amplitude
for a longitudinally polarized initial photon.
 The parameter integral over $v$ gives the function  
\mbox{$ [ \ln ( x + \xi - i0) - \ln (2 \xi-i0) ]/(x-\xi)$}  
which  is 
regular at $x = \xi$ and has a 
logarithmic singularity at $x = - \xi$.
The integral over $x$ exists if  
$
R(x, \xi) $ 
is bounded at $x=-\xi$, which again is the case
in the DD-based models described in  Ref. \cite{Musatov2000}.  
The third term of Eq. (\ref{Compton_Anikin}) 
corresponds to the transverse polarization  of the initial photon. 
In this case,  one faces  the integrand   
$1/[v(x-\xi)+i0]$  which produces $dv/v$   divergence for the $v$-integral
at  the lower limit. One may  hope to  get a finite result 
only if the integral 
\beq
I(\xi) \equiv \int_{-1}^1 dx
  \frac{R(x,\xi)}{x - \xi + i0} \label{Ixi} 
\eeq
vanishes. 
From the definition of the skewed 
distributions $H(x, \xi)$ and  $ A (x, \xi)$ (\ref{reduction}) 
it  follows that  
 $$\frac{{\partial} A (x, \xi)}{\partial x} =
-\frac{ {\partial} H (x, \xi)}{\partial \xi} .$$
Hence, one can substitute 
  $R (x, \xi) $ 
by the combination  ${\partial} H (x, \xi)/{\partial x} + 
{\partial} H (x, \xi)/{\partial \xi} $ similar to that used in
\cite{Kivel:2000rb}  within the context of 
WW approximation 
(in Refs. \cite{Belitsky:2000vx,Kivel:2000rb},  full SPD's 
 are implied
 while our $ H (x, \xi)$ does not include the PW term).   
 By analogy with  Ref.  \cite{Kivel:2000rb}, we integrate the
 $ {\partial} H (x, \xi)/{\partial x}$ term by parts.
 This   gives 
\be
I(\xi) = \frac{d}{d \xi} 
\int_{-1}^1 dx \frac{H(x, \xi)}{x - \xi + i0} \ ,
\label{xider}  
\ee
i.e., the $\xi$ derivative of the  twist-2 
contribution. In general, the latter has a  nontrivial 
$\xi$-dependent form  determined 
by the shape of SPDs (see, however, the discussion of the
PW contribution below).  
We conclude that  the twist-3 part of the tensor 
amplitude $T_{\mu\nu}$ diverges in case of 
the transverse polarization of the initial photon. A similar 
observation has been  recently made in Ref.\cite{Kivel:2000rb}. 
 However, it is easy to see that the relevant 
 tensor structure $\Delta_\mu (q_\nu +\xi p_\nu)$
 is just a truncated version of the exactly  gauge 
 invariant combination
 $\Delta_\mu {q_2}_{\nu}$ which 
 has zero projection onto
 the polarization vector 
 $\epsilon_2^\nu$ of the final real  photon: $(\epsilon_2 q_2)=0$. 
 The structure $\Delta_\mu {q_2}_{\nu}$ is obtained if one uses 
 the original full form of the DD parametrization
 (\ref{para5}).  
 It appears  from the term with   the   exponential
  factor of the argument 
  $  -i [v(\tilde x +\xi \alpha)(pz)+
  \bar v \xi (pz) + (v \alpha +\bar v) (\Delta z)/2] $
  which is 
  obtained by combining 
  the   sine/cosine functions  and the  
  exponential factor  in the second
  term of Eq. (\ref{para5}). In the Compton amplitude, 
  it gives rise to a  contribution in which the argument 
  of the quark propagator is
  $q+v(\tilde x +\xi \alpha)p  + \bar v \xi p + (v \alpha + \bar v)\Delta/2$.
   Since $(\Delta q)$,   $(\Delta p$) and $\Delta^2 $ are negligible, 
     the denominator factors
  in Eq. (\ref{Compton_Anikin}) remain unchanged. 
 In  numerators,  representing $(v \alpha + \bar v)\Delta/2$
 as $[1-(1-\alpha)v]\Delta/2$,  we observe that 
   $\Delta_{\mu} (q_\nu+\xi p_\nu)$  
      converts into the 
 $\Delta_{\mu} (q_\nu+\xi p_\nu +\Delta_\nu/2 )= \Delta_\mu {q_2}_{\nu}$ 
 term  
 plus a  $v \Delta_{\mu} \Delta_{\nu}$  type 
 contribution corresponding to a new SPD built  from
 the $(1-\alpha)^2 f(\tilde x, \alpha)$ DD (cf. (\ref{reduction})). 
 Due to the  extra $v$ factor, the $v$-integral for 
 the latter contribution is finite. 
   Hence, for the physical DVCS amplitude, 
 we find  no evidence for factorization breaking
 in the  kinematical twist--3 contributions, both in their
$1/\sqrt{-q^2}$ and $1/{q^2}$ terms.
It is quite possible that factorization breaks down 
at the $1/{q^2}$ level,  
but one needs to analyze $O(z^2)$  suppressed 
terms (i.e., twist--4 contributions) to see if it really happens.  

 {\it WW-type representation.}
We can express our  results  
 in 
another  form, introducing new skewed distributions  related to 
$R(x,\xi)$  $via$  an integral  transformation 
similar to that used by Wandzura and Wilczek \cite{WW}  
 \footnote{WW-type  integrals  of parton distributions
     have 
originally  appeared 
within the   
$\xi$-scaling
formalism \cite{Georgi:1976ve} .} .
 This allows us to establish the equivalence 
between  our approach and the WW-type  approximation 
proposed in Ref.\cite{Belitsky:2000vx}. 
Treating the combination $xv+\bar v \xi$ in 
(\ref{Compton_Anikin}) as a new 
variable  we  define 
\begin{eqnarray}  
R_W (x, \xi) \equiv   
\int_{-1}^1 R (y, \xi) \, dy \int_0^1   
\, \delta ( yv + \bar v \xi -  x)   \, dv   
\, =  \, \, \theta ( x > \xi) \int_{x}^1
\frac{R(y,\xi)}{y-\xi}\, dy  -  \, \theta ( x < \xi) \int_{-1}^x 
\frac{R(y,\xi)}{y-\xi}\, dy  \  . 
 \label{rw} 
 \end{eqnarray} 
In terms of this transform, the matrix element of the 
vector operator (\ref{para7}) 
can be expressed as 
\begin{eqnarray}  
\frac12 \langle \, p-r/2 \, |\,  
{\cal O}_{\sigma} (z\, |\, 0) \, | \, p+r/2 \rangle  
&=&
  \int_{-1}^1 d x  \, e^{-i x(pz) }
\biggl \{ 
p_{\sigma}   H(x, \xi) 
- \frac{i}{2} \, p_{\sigma} \, 
 (\Delta z)\,  A(x, \xi)  \nonumber \\ &  + &  
\frac{1}{4} \, \biggl (\Delta_{\sigma}   - 
p_{\sigma}\frac{(\Delta z)}{(pz)}\biggr )
\biggl [ 
R_W (x, \xi)- 
 R_W (-x, \xi)  
 \biggr ] \biggr \}   \ .   
\label{para8} 
\end{eqnarray}
Note that only the odd part of $R_W (x, \xi)$ contributes 
here. In case of the axial operator (\ref{parax5})
\begin{eqnarray} 
\frac12 \langle \, p-r/2 \, |\,  {\cal O}_{5 \sigma} 
(z\, |\, 0) \, | \, p+r/2 \rangle 
& = &   \frac{i}{4} \,   \epsilon_{\sigma\alpha\beta\gamma} \,
\frac{ z_\alpha}{(pz)} 
 \Delta_{\beta}    p_{\gamma}
\int_{-1}^1 d x \,e^{-i  x (pz) } 
 \biggl [ R_W (x,\xi) + R_W (-x,\xi) \biggr ]  
\label{parax6} 
\end{eqnarray}
only the even   part of $R_W (x, \xi)$ appears. 
The  part  
 of the 
Compton amplitude (\ref{Compton_Anikin})  
containing  $R(x,\xi)$  
can   be written  in terms of this new
function as  
\begin{eqnarray} 
  \frac1{2(qp)}
\int_{-1}^1 
\, 
\biggl [ \frac{\Delta^{\mu} ( q^{\nu} + \xi  p^{\nu})}
{x-\xi +i0} + \frac{( q^{\mu} + 3 \xi  p^{\mu})\Delta^{\nu} }
{x+\xi -i0 }\biggr ]
R_W(x,\xi) \, dx \  . 
 \label{comfin} 
\end{eqnarray}

The integrals with $1/(x \pm \xi \mp i0)$  converge
only if the function $ R_W(x,\xi)$
is continuous for $x=\pm \xi$. According to Eq. (\ref{rw}),  
$ R_W(x,\xi)$  is  given by the integral
of $R(y,\xi)/(y-\xi)$ from $x$ to 1 if
$x> \xi$ and from  $x$ to $-1$ if $x < \xi$.
Evidently,  $x=-\xi$ is not a special 
point in the integral transformation (\ref{rw}), hence 
the function $ R_W(x,\xi)$ 
is continuous at $x=-\xi$. 
However, it  is extremely unlikely 
 that the limiting values approached by 
 $ R_W(x,\xi)$ for $x=\xi$  
from below and from above do coincide. 
Indeed, the difference of the two limits 
can be written as the principal value integral  
(compare with \cite{Kivel:2000rb}) 
\begin{eqnarray}
 R_W(\xi+0, \, \xi) -  R_W(\xi-0,\, \xi )
 = {\rm P} \int_{-1}^1 \frac{R(y,\xi)}{y-\xi}\, dy \ , 
 \label{pv} 
 \end{eqnarray}
 which  can be converted  into the $\xi$-derivative
 of the real part of the twist--2 contribution. 
 This means that  the singularity,  
which we observed as a  straight divergence of the $dv/v$ 
integral, in this approach 
appears due to an unavoidable  discontinuity  of the 
$R_W (x, \xi)$ transform   
at $x = \xi$.

{\it Contribution from the PW--term.} 
The contribution of the PW term to the 
vector operator 
\begin{eqnarray} 
\frac{1}{2} \, 
\langle p - r/2 \, | \, {\cal O}_{\sigma}(0\,|\,z)  \,  
| \, p + r/2 \rangle_{\mbox{\scriptsize PW--Term}}
\;\; = \;\;
\frac{r_{\sigma}}{2}  \int_{-1}^1 
d \alpha  \, e^{-i \alpha (rz)/2} \, 
D (\alpha) 
\label{para_PW} 
\end{eqnarray}
 has a simple structure corresponding to a 
parton picture in which the partons carry the fractions $(1 \pm \alpha)/2$ 
of the momentum transfer $r$. Since only one momentum  $r$  is 
involved, this term can contribute only to the totally symmetric part
of the vector string operator: it  ``decouples'' in the 
reduction relations (\ref{rest_total_derivative}). In particular, the  
{\rm PW} term  does not contribute 
to the second contribution  in Eq. (\ref{string_deconstructed_scalar})
which is generated by decomposition
of the axial string operator:  both derivatives, with respect to  
$X$ and $z$, give rise to the momentum transfer $r$,  whence the 
contraction with the $\epsilon$--tensor in 
(\ref{string_deconstructed_scalar}) gives zero.
Thus, the PW-contribution  should be transverse by 
itself. Indeed, a straightforward calculation gives 
\begin{equation}
T_{\mu\nu}|_{\mbox{\scriptsize PW}}  \;\; = \;\; 
\frac{1}{(rq)}
\biggl [ r_{\mu} q_{\nu} +  q_{\mu} r_{\nu}  -  g_{\mu \nu} (rq)
+ r_{\mu} r_{\nu} \biggr ] \; \int_{-1}^1 
\frac{D (\alpha)}{\alpha-1}  \,  d \alpha  \  , 
\label{PW} 
\end{equation} 
which evidently 
satisfies  
$
q_{\mu} T_{\mu \nu}|_{\mbox{\scriptsize PW}}  =  0, \ 
r_{\mu} T_{\mu \nu}|_{\mbox{\scriptsize PW}}  = 0.
$
Hence,  this term can be treated as a separate contribution. 

Alternatively, one may include it into the basic SPD $H(x,\xi)$ 
and all SPD's derived from $H(x,\xi)$.
Specifically, for $\xi >0$, the PW contribution 
to $H(x,\xi)$ is $D (x/\xi) \, \theta (|x| \leq \xi)$ 
\cite{PW}; it contributes   $(\xi-x)D'(x/\xi)\, \theta (|x| \leq \xi)/\xi^2$
[where $D'(\alpha) \equiv (d/d\alpha) D(\alpha)] $ to 
$R(x,\xi)$; 
furthermore,  the  PW contribution to  $R_W(x,\xi)$ is 
$D (x/\xi)\, \theta (|x| \leq \xi)/\xi$. 
Inserting these functions into Eqs. (\ref{Compton_Anikin}) and 
(\ref{comfin})
one rederives Eq. (\ref{PW}). 
One can also observe that the PW term gives zero contribution
into $I(\xi)$, Eq. (\ref{Ixi}).

\section{Conclusions}
\label{sec_conclusions}
In this paper,  we have studied the DVCS amplitude $T_{\mu\nu}$ 
making use of the light--cone expansion in terms of QCD string 
operators in the coordinate space. We have demonstrated  that 
transversality of the light--cone expansion
can be maintained by including  
a minimal set of ``kinematical'' 
twist--3 operators, which appear as total 
derivatives of twist--2 operators. 
Incorporating the formalism of double 
distributions, we  established that the kinematical
twist--3 contributions are described by 
a   skewed parton distribution $R_W(x,\xi)$
which can be derived from 
the basic twist--2 double distribution 
$f(\tilde x, \alpha)$.
The new SPD $R_W(x,\xi)$ 
has the structure of a generalized  WW  transform. 
Due to discontinuities of $R_W(x,\xi)$ at  $x=\xi$,
the factorization for the tensor amplitude $T_{\mu\nu}$
 breaks down at the twist-3 level
for the part  corresponding 
to the transverse virtual photon. 
 However, this term gives zero projection
 onto the polarization vector of the 
 final real photon.

\section{Acknowledgements}
We are grateful to I.I.~Balitsky, J.~Bl\"umlein, V.M.~Braun, 
B.~Geyer, N.~Kivel, M.~Lazar, I.V.~Musatov, M.V.~Polyakov  and 
O.V.~Teryaev for many stimulating discussions. 
We thank M. Diehl for useful comments.
This work was supported by the US 
 Department of Energy under contract
DE-AC05-84ER40150,   by the Deutsche 
Forschungsgemeinschaft (DFG), by
the German Ministry of Education and Research (BMBF),
and by COSY, J\"ulich.
A.R. expresses his deep gratitude
to  K. Goeke for  warm hospitality 
in  Bochum and support.

\end{document}